\documentclass[showpacs,10pt,twocolumn,prl]{revtex4-1}

\usepackage{amsmath}
\usepackage{amssymb}
\usepackage{graphicx}
\usepackage{amssymb}
\usepackage{graphics}
\usepackage{epsfig}
\usepackage{CJK}
\usepackage{color}
\usepackage{soul}

\begin{document}

\begin{CJK*}{GBK}{Song}
\title{Electrical and thermal transport in van der Waals magnets 2H-M$_x$TaS$_2$ (M = Mn, Co)}
\author{Yu Liu,$^{1,2,\dag}$ Zhixiang Hu,$^{1,3}$ Xiao Tong,$^{4}$ Eric D. Bauer,$^{2}$ and C. Petrovic$^{1,3,\ddag}$}
\affiliation{$^{1}$Condensed Matter Physics and Materials Science Department, Brookhaven National Laboratory, Upton, New York 11973, USA\\
$^{2}$Los Alamos National Laboratory, Los Alamos, New Mexico 87545, USA\\
$^{3}$Materials Science and Chemical Engineering Department, Stony Brook University, Stony Brook, New York 11790, USA\\
$^{4}$Center for Functional Nanomaterials, Brookhaven National Laboratory, Upton, New York 11973, USA}
\date{\today}

\begin{abstract}
  We report a detailed study of electrical and thermal transport properties in 2H-M$_x$TaS$_2$ (M = Mn, Co) magnets where M atoms are intercalated in the van der Waals gap. The intercalation induces ferromagentism with an easy-plane anisotropy in 2H-Mn$_x$TaS$_2$, but ferromagnetism with a strong uniaxial anisotropy in 2H-Co$_{0.22}$TaS$_2$, which finally evolves into a three-dimensional antiferromagnetism in 2H-Co$_{0.34}$TaS$_2$. Temperature-dependent electrical resistivity shows metallic behavior for all samples. Thermopower is negative in the whole temperature range for 2H-Co$_x$TaS$_2$, whereas the sign changes from negative to positive with increasing Mn for 2H-Mn$_x$TaS$_2$. The diffusive thermoelectric response dominates in both high- and low-temperature ranges for all samples. A clear kink in electrical resistivity, a weak anomaly in thermal conductivity, as well as a slope change in thermopower were observed at the magnetic transitions for 2H-Mn$_{0.28}$TaS$_2$ ($T_\textrm{c}$ $\approx$ 82 K) and 2H-Co$_{0.34}$TaS$_2$ ($T_\textrm{N}$ $\approx$ 36 K), respectively, albeit weaker for lower $x$ crystals. Co-intercalation promoted ferromagnetic to antiferromagnetic transition is further confirmed by the Hall resistivity; the sign change of the ordinary Hall coefficient indicates a multi-band behavior in 2H-Co$_x$TaS$_2$.
\end{abstract}
\maketitle
\end{CJK*}

\section{INTRODUCTION}

Recent discovery of intrinsic long-range magnetic order in ultrathin crystals of two-dimensional (2D) van der Waals (vdW) magnets, for instance, FePS$_3$, Cr$_2$Ge$_2$Te$_6$, CrI$_3$, Fe$_3$GeTe$_2$, VSe$_2$, and MnSe$_2$, has motivated significant number of studies devoted to its physical mechanism and tuning of functionalities in vdW heterostructures and devices \cite{Geim,Novoselov,Lee,Huang,Gong,Deng,Bonilla,Hara}.

Intercalated transition metal dichalcogenides commonly feature 3$d$ atoms in vdW gap and exhibit diverse magnetic properties \cite{Parkin,Friend0}. For example, 2H-M$_{1/3}$TaS$_2$ with M = V, Cr, Mn is ferromagnetic (FM); 2H-Fe$_x$TaS$_2$ is FM for 0.2 $\leq$ $x$ $\leq$ 0.4 but is antiferromagnetic (AFM) for higher $x$; whereas 2H-M$_{1/3}$TaS$_2$ with M = Co, Ni is AFM \cite{Parkin}.  2H-Mn$_x$TaS$_2$ is a soft FM with an easy-plane anisotropy \cite{Hinode,Onuki,Cussen,Li,Shand,Zhang1,YL1}; its anisotropic magnetoresistance (MR) indicates a possible field-induced novel magnetic structure \cite{Zhang1,YL1}. The most widely studied member in this family is 2H-Fe$_{1/4}$TaS$_2$, which exhibits FM with a strong uniaxial anisotropy, and shows large magnetocrystalline anisotropy and MR, sharp switching in magnetization, and anomalous Hall effect \cite{Morosan,Hardy,Chen,Cai,Zhang}. Recently discovered 2H-Co$_{0.22}$TaS$_2$ also shows FM with a strong uniaxial anisotropy \cite{YL}, in contrast to 2H-Co$_{1/3}$TaS$_2$ exhibiting AFM \cite{VanLaar}. As we can see, magnetic order evolves from FM in M = V, Cr, Mn through FM and AFM in M = Fe, Co to AFM in M = Ni for 2H-M$_x$TaS$_2$. To unveil the physical origin of magnetic order evolution, a detailed study on 2H-M$_x$TaS$_2$ with intermediate Mn, Fe, Co intercalations will be helpful. In addition to its magnetism, the study of transport properties of 2H-M$_x$TaS$_2$ are also required for further spintronic applications.

In this work we fabricated a series of 2H-M$_x$TaS$_2$ (M = Mn, Co) single crystals, and systematically studied their magnetic, electrical and thermal transport properties. Since few-layer graphene/2H-TaS$_2$ heterostructures preserve 2D Dirac states with robust spin-helical structure of interest for spin-logic circuits \cite{LiLijun}, the possibility of integration of robust magnetism is of high interest for spintronic and calls for nanofabrication of graphene/2H-M$_x$TaS$_2$ (M = Mn, Co) heterostructures and devices.

\section{EXPERIMENTAL DETAILS}

Single crystals of 2H-M$_x$TaS$_2$ (M = Mn, Co) with typical hexagonal shape were grown by chemical vapor transport method with iodine agent. The raw materials of Mn, Co, Ta, and S powders were sealed in an evacuated quartz tube, and then heated for a week in a two-zone furnace with the source temperature of 1000 $^\circ$C and the growth temperature of 900 $^\circ$C. The average stoichiometry was determined by examination of multiple points on cleaved fresh surfaces and checked by multiple samples from the same batch using energy-dispersive x-ray spectroscopy in a JEOL LSM-6500 scanning electron microscope. X-ray diffraction (XRD) data were acquired on a Rigaku Miniflex powder diffractometer with Cu $K_{\alpha}$ ($\lambda=0.15418$ nm). Scanning tunneling microscope (STM) was carried out using Scienta Omicron VT STM XA 650 with Matrix SPM Control System in an UHV chamber with 2$\times$10$^{-10}$ torr base pressure. STM topography is obtained in the constant current mode with positive sample bias (unoccupied state image) at room temperature. The SPIP software (Image Metrology, Denmark) was used to process and analyze STM images. Crystal was cleaved by scotch tape to obtain a fresh surface and then transferred from air into UHV-STM chamber without any in-site cleaning treatment. The magnetization was measured in quantum design MPMS-XL5 system. Electrical and thermal transport were measured by standard four-probe method in quantum design PPMS-9 system. In order to effectively eliminate the longitudinal resistivity contribution due to voltage probe misalignment, Hall resistivity was calculated by the difference of transverse resistance measured at positive and negative fields, i.e., $\rho_{xy} = [\rho(\textrm{+H})-\rho(\textrm{-H})]/2$.

\section{RESULTS AND DISCUSSIONS}

Figure 1(a,b) presents the crystal structure with space group $P6_322$ of 2H-M$_x$TaS$_2$ (M = Mn, Co) from the side and the top views, respectively. The sharp peaks in the XRD 2$\theta$ scans can be indexed with (00l) planes [Fig. 1(c)], indicating that the plate surface of single crystal is normal to the $\mathbf{c}$-axis. With increasing $x$, the (00l) peaks gradually shift to lower angles, as clearly seen in the enlarged (002) peak, indicating an expansion of lattice parameter $c$. The values of $c$ can be extracted by using the Bragg's law; monotonically increases with $x$, confirming that the M atoms are intercalated and expand the vdW gap of 2H-TaS$_2$. Figure 1(d) shows a STM topography of 2H-Mn$_{0.28}$TaS$_2$ crystal surface, from which a triangular lattice can be observed. Figure 1(e) shows the Laue x-ray diffraction pattern of 2H-Co$_{0.34}$TaS$_2$ crystal, confirming the six-fold symmetry of the hexagon structure and well orientation along (00l) direction.

\begin{figure}
\centerline{\includegraphics[scale=1]{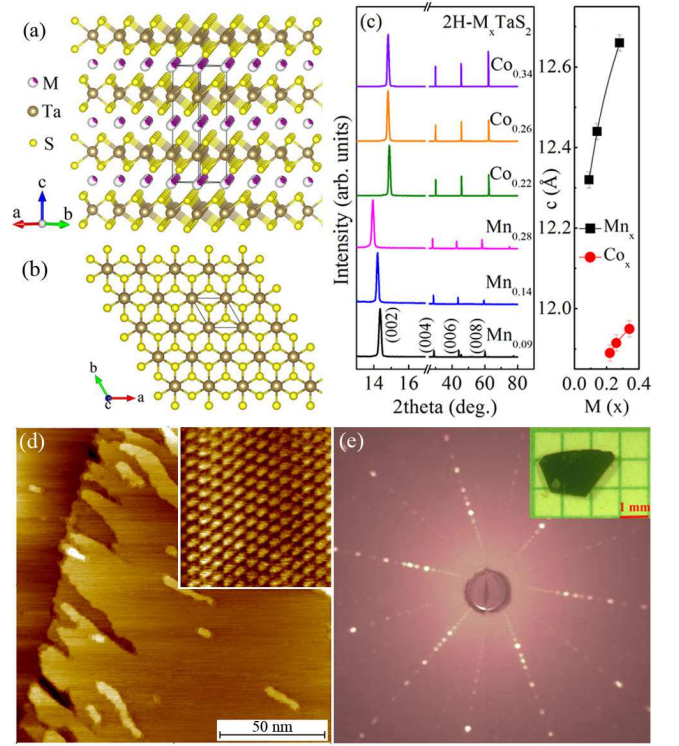}}
\caption{(Color online) Crystal structure of 2H-M$_x$TaS$_2$ (M = Mn, Co) shown from the (a) side view and (b) top view, respectively. (c) Single crystal x-ray diffraction patterns of 2H-M$_x$TaS$_2$ (M = Mn, Co) and the evolution of lattice parameter $c$ with intercalation content $x$. (d) Scanning tunneling microscope topography of the sample surface for 2H-Mn$_{0.28}$TaS$_2$. (e) Laue x-ray pattern on the shown surface of 2H-Co$_{0.34}$TaS$_2$ crystal with the six-fold symmetry of the hexagon structure.}
\label{XRD}
\end{figure}

\begin{table*}[!htb]
\caption{\label{tab}The actual chemical composition, lattice parameter $c$, the ratio of $\chi_{\textrm{ab}}/\chi_\textrm{c}$ at $T$ = 2 K
  in field cooling curves, and the parameters obtained from the Curie-Weiss fit of the $1/\chi$ vs $T$ data and isothermals at $T$ = 2 K for 2H-M$_x$TaS$_2$ (M = Mn, Co) single crystals. The values of $T_\textrm{c}$ and $T_\textrm{N}$ are determined by the minima of the $d\chi/dT$ curves and the maxima of the $\chi(T)$ curves in field cooling mode along the easy $\mathbf{ab}$-plane for Mn and the easy $\mathbf{c}$-axis for Co, respectively. The RWR represents the Rhodes-Wohlfarth ratio.}
\begin{ruledtabular}
\begin{tabular}{llllllllllll}
         & $x$ & $c$ & $\chi_{\textrm{ab}}/\chi_\textrm{c}$ & Field & $T_\textrm{c}$ & $T_\textrm{N}$ & $\theta$ & $C$ & $P_{\textrm{eff}}$ & $P_{\textrm{sat}}$ & RWR \\
         & & ({\AA}) & (2 K) & & (K) & (K) & (K) & (K emu mol$^{-1}$ Oe$^{-1}$) & ($\mu_\textrm{B}/\textrm{M}$) & ($\mu_\textrm{B}/\textrm{M}$) & \\
  \hline
  M = Mn & 0.09 & 12.32(2) & 4.03 & \textbf{H}$\parallel$\textbf{ab} & 11 && 8(1) & 0.369(2) & 5.73(4) & 1.15(7) & 4.2 \\
         & 0.14 & 12.44(2) & 9.71 & \textbf{H}$\parallel$\textbf{ab} & 22 && 47(1) & 0.541(2) & 5.56(2) & 2.76(3) & 1.7 \\
         & 0.28 & 12.66(2) & 6.06 & \textbf{H}$\parallel$\textbf{ab} & 82 && 101(1) & 1.01(1) & 5.37(3) &4.40(1) & 1.0 \\
  \hline
  M = Co & 0.22 & 11.89(2) & 0.04 & \textbf{H}$\parallel$\textbf{c} & 26 && 39(2) & 0.164(5) & 2.4(2) & 0.60(1) & 2.7 \\
         & 0.26 & 11.91(2) & 0.65 & \textbf{H}$\parallel$\textbf{c} && 26 & 6(1) & 0.232(3) & 2.7(1) & \\
         & 0.34 & 11.95(2) & 1.08 & \textbf{H}$\parallel$\textbf{c} && 36 & -88(9) & 0.38(3) & 3.0(1) &
\end{tabular}
\end{ruledtabular}
\end{table*}

\begin{figure*}
\centerline{\includegraphics[scale=1]{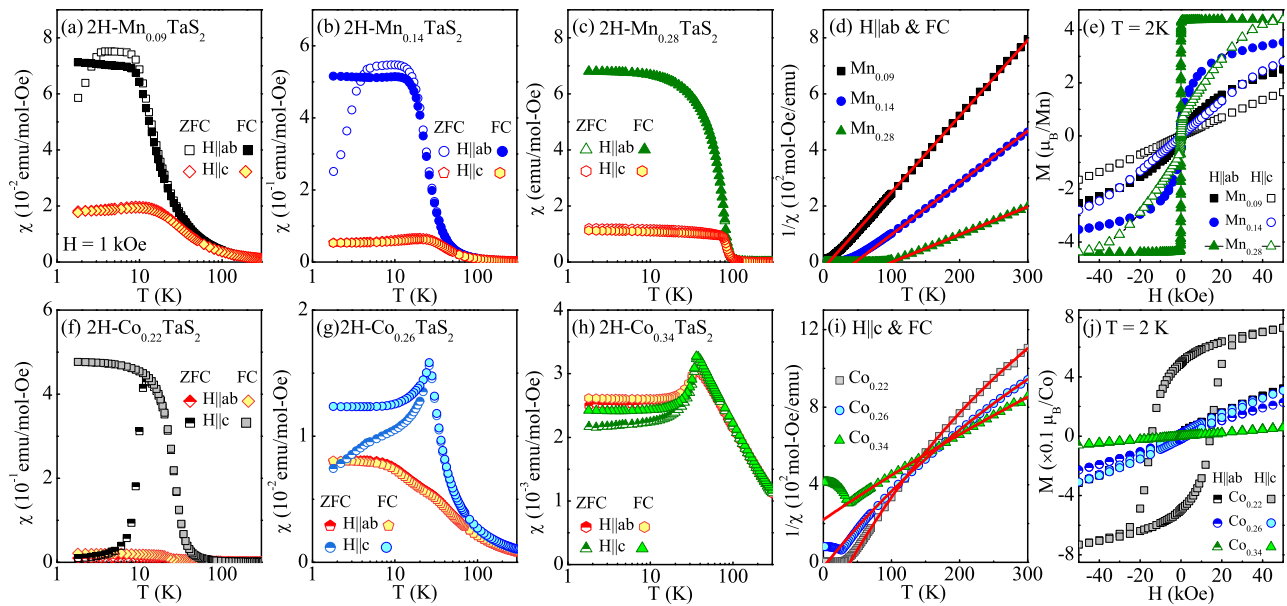}}
\caption{(Color online) (a-c) Temperature-dependent magnetic susceptibility $\chi$(T) in both zero-field cooling (ZFC) and field cooling (FC) modes with $\mathbf{H} \parallel \mathbf{ab}$-plane and $\mathbf{H} \parallel \mathbf{c}$-axis at $H$ = 1 kOe for 2H-Mn$_x$TaS$_2$. (d) The inverse susceptibility $1/\chi$(T) with $\mathbf{H} \parallel \mathbf{ab}$-plane fitted by the Curie-Weiss law (solid lines) for 2H-Mn$_x$TaS$_2$. (e) Field-dependent magnetization with both $\mathbf{H} \parallel \mathbf{ab}$-plane and $\mathbf{H} \parallel \mathbf{c}$-axis at $T$ = 2 K for 2H-Mn$_x$TaS$_2$. (f-h) Temperature-dependent magnetic susceptibility $\chi$(T) in both ZFC and FC modes with $\mathbf{H} \parallel \mathbf{ab}$-plane and $\mathbf{H} \parallel \mathbf{c}$-axis at $H$ = 1 kOe for 2H-Co$_x$TaS$_2$. (i) The inverse susceptibility $1/\chi$(T) with $\mathbf{H} \parallel \mathbf{c}$-axis fitted by the Curie-Weiss law (solid lines) for 2H-Co$_x$TaS$_2$. (j) Field-dependent magnetization with both $\mathbf{H} \parallel \mathbf{ab}$-plane and $\mathbf{H} \parallel \mathbf{c}$-axis at $T$ = 2 K for 2H-Co$_x$TaS$_2$.}
\label{XRD}
\end{figure*}

Figure 2(a-c) exhibits the temperature dependence of magnetic susceptibility $\chi$(T) measured at $H$ = 1 kOe applied in the $\mathbf{ab}$-plane and along the $\mathbf{c}$-axis with zero-field cooling (ZFC) and field cooling (FC) modes for 2H-Mn$_x$TaS$_2$. A sharp upturn in $\chi$(T) was observed with $\mathbf{H} \parallel \mathbf{ab}$ as temperature decreases, suggesting a paramagnetic(PM)-FM transition. The transition temperatures $T_\textrm{c}$ are defined by the minima in $d\chi/dT$ and listed in Table 1. The low-temperature values of $\chi$(T) for $\mathbf{H} \parallel \mathbf{ab}$ are larger than those for $\mathbf{H} \parallel \mathbf{c}$ for 2H-Mn$_x$TaS$_2$, indicating the magnetic moments of Mn tend to be arranged in the $\mathbf{ab}$-plane with an easy-plane anisotropy. With further temperature decreasing, the bifurcation between ZFC and FC curves below 4 K for low Mn content [Fig. 2(a,b)] is due to a possible spin glass state, which was previously investigated by the ac susceptibility measurement, and is attributed to the inhomogeneity of Mn-intercalation \cite{Li,Zhang1}. The freezing of spin glass may also contribute to the weak increase of resistivity (see below). Between the freezing temperature and $T_c$, the ZFC values are larger than the FC values for low Mn content [Fig. 2(a,b)], indicating a possible large magnetostriction. This phenomenon was also observed in the phase separated manganite \cite{Zhao}. The inverse susceptibility $1/\chi$ from 100 to 300 K can be well fitted by the Curie-Weiss law $\chi = \chi_0 + C/(T-\theta)$ [Fig. 2(d)], where $\chi_0$ is a temperature-independent term, $C$ and $\theta$ are the Curie-Weiss constant and Weiss temperature, respectively. The derived $\theta$ is positive and increases with increasing Mn content [Table I], indicating dominance of FM exchange interactions. Further increasing Mn from $x \sim$ 0.25 to 0.5 \cite{Hinode}, the value of $\theta$ will decreases and be negative at $x$ = 0.5; i.e., the magnetic interactions change from FM to AFM in Mn$_{0.5}$TaS$_2$. This may originate from the shorter Mn-Mn distance in the vdW gap plane with the increase of Mn content \cite{Bongers}. The derived effective moment $P_{\textrm{eff}}$ (= $\sqrt{8C/x}$) decreases from 5.73(3) to 5.37(4) $\mu_\textrm{B}$/Mn [Table I]. The valence states of Mn are estimated to be divalent (spin only moment of 5.92 $\mu_\textrm{B}$ for Mn$^{2+}$). Although the $P_{\textrm{eff}}$ value in the samples with $x$ = 0.4 $\sim$ 0.5 decreases to 5.0 $\mu_\textrm{B}$/Mn, which is close to the spin-only Mn$^{3+}$ value of 4.9 $\mu_\textrm{B}$ \cite{Hinode}; the electron spin resonance measurement gives the signal corresponding to the Mn$^{2+}$ ion \cite{Hinode}. Polarized neutron study has confirmed that the moment on Mn sites depressed by about 15\% compared with the expected value for Mn$^{2+}$ in Mn$_{0.25}$TaS$_2$ \cite{Parkin0}. Then we estimated the Rhodes-Wohlfarth ratio (RWR) for 2H-Mn$_x$TaS$_2$, which is defined as $P_\textrm{c}/P_{\textrm{sat}}$ with $P_\textrm{c}$ calculated from $P_\textrm{c}(P_\textrm{c}+2) = P_{\textrm{eff}}^2$; and the $P_{\textrm{sat}}$ is the saturation moment estimated by using a linear fit of $M$(H) above a magnetic field of 3 T \cite{Wohlfarth,Moriya}. The value of RWR [Table I] decreases with increasing Mn content, from 4.2 for 2H-Mn$_{0.09}$TaS$_2$ to 1.0 for 2H-Mn$_{0.28}$TaS$_2$, indicating a gradual evolution from itinerant to localized character. Although the moment is principally localized on Mn sites, there is a significant spin polarization of the conduction electrons. The 3$d$-electron-conduction-electron interaction will result in some loss of magnetic moment, accounting for our observation of the decreased effective moment as Mn content increases. Field-dependent magnetization at $T$ = 2 K [Fig. 2(e)] confirms the easy-plane anisotropy in 2H-Mn$_x$TaS$_2$, and an enhanced saturation moment for higher $x$. Negligible hysteresis loop (coercive field $H_\textrm{c}$ $<$ 10 Oe) indicates a soft in-plane FM character.

Figure 2(f) shows the temperature dependence of $\chi$(T) for 2H-Co$_{0.22}$TaS$_2$, suggesting a FM ground state with strong uniaxial anisotropy, similar to 2H-Fe$_{1/4}$TaS$_2$ \cite{Morosan,Hardy,Chen}. The different tendency between ZFC and FC curves at low temperature is due to the magnetic domain creep effect \cite{YL}, which is expected for long-range FM state with magnetic anisotropy (hard ferromagnet and sizable coercive field) and/or multidomain structure \cite{Hechang}. The FM exchange interaction becomes weaker with increasing Co content, as well as the parameter of magnetic anisotropy estimated by the value of $\chi_{\textrm{ab}}/\chi_\textrm{c}$ at $T$ = 2 K [Table I]. Then 2H-Co$_{0.26}$TaS$_2$ features an AFM-dominated peak in $\chi$(T) when $\mathbf{H} \parallel \mathbf{c}$, while it shows FM with a two-step feature when $\mathbf{H} \parallel \mathbf{ab}$, indicating competed FM and AFM interactions in 2H-Co$_{0.26}$TaS$_2$. An almost isotropic three-dimensional (3D) AFM order finally dominates in 2H-Co$_{0.34}$TaS$_2$ [Fig. 2(h)], in line with previous reports \cite{VanLaar,Parkin1}, where the transition temperature $T_\textrm{N}$ can be defined as the temperature of the maximum in $\chi$(T). The Weiss temperature $\theta$ evolves from positive 39(2) K for 2H-Co$_{0.22}$TaS$_2$ to negative -88(9) K for 2H-Co$_{0.34}$TaS$_2$ [Table I]. The derived $P_{\textrm{eff}}$ of 2.4(2) $\sim$ 3.0(1) $\mu_\textrm{B}$/Co is smaller than the spin-only moment of 3.87 $\mu_\textrm{B}$ for Co$^{2+}$. Here we propose to attribute the similar loss of Co magnetic moment to a 3$d$-electron-conduction-electron mixing interaction. This was also observed in Co$_{0.33}$NbS$_2$ by an unpolarized neutron single-crystal study \cite{Parkin1}. Figure 2(j) shows the field-dependent magnetization at 2 K for 2H-Co$_x$TaS$_2$, where a large coercivity $H_\textrm{c}$ of 14.3 kOe with $\mathbf{H} \parallel \mathbf{c}$ was observed in 2H-Co$_{0.22}$TaS$_2$ \cite{YL}.

In 2H-M$_x$TaS$_2$ charge transfer from $3d$ transition metal to the Ta $d$ band is evident; the remaining $d$ electrons on the M ions are localized and show magnetic moments. The remarkable difference in 2H-M$_x$TaS$_2$ with intermediate Mn, Fe, Co intercalations is the magnetic anisotropy; i.e., FM with an easy-plane anisotropy for 2H-Mn$_x$TaS$_2$ \cite{YL1}, however an easy-$\textbf{c}$-axis anisotropy for 2H-Fe$_x$TaS$_2$ \cite{Choe} and 2H-Co$_{0.22}$TaS$_2$ \cite{YL}. As we know, the two main interactions responsible for the magnetic anisotropy are the single-ion anisotropy and the dipolar anisotropy. In 2H-Mn$_x$TaS$_2$ the Mn 3$d\downarrow$ band is unoccupied, while the Mn 3$d\uparrow$ is nearly completed occupied. The extra holes in the 3$d\uparrow$ occupy $e_g^\sigma$-type state, which have no orbital momentum. Therefore the magnetic anisotropy in 2H-Mn$_x$TaS$_2$ is dominated by dipolar interactions, leading to an easy-plane anisotropy \cite{Dijkstra}. However, in 2H-Fe$_x$TaS$_2$ it is a single-ion effect due to interaction of spin with orbital moment of the partially occupied Fe 3$d\downarrow$ band. As a result of the trigonal distortion of the octahedra of S atoms, the lowest-energetic $e_g^\pi$ states (derived from $t_{2g}$ states without distortion) have an orbital momentum parallel the $\textbf{c}$-axis, and spin-orbital interaction leads to an easy-$\textbf{c}$-axis anisotropy. We propose the similar mechanism for FM in 2H-Co$_{0.22}$TaS$_2$ due to the large orbital contributions.

Considering the anomalies in the electrical resistivity and the thermopower at the ordering temperatures (see below), the conduction electrons also play an important role in mediating the exchange interactions between the magnetic moments, i.e., the Ruderman-Kittel-Kasuya-Yosida (RKKY) interaction due to the high carrier density in 2H-TaS$_2$ \cite{Aristov,Yosida,Fukuma,YamasakiY,NiuJ}. With higher intercalation in 2H-Mn$_{0.5}$TaS$_2$, 2H-Fe$_{0.4}$TaS$_2$, and 2H-Co$_{0.33}$TaS$_2$ \cite{Hinode,Narita,VanLaar}, the nearest-neighboring M atoms will be closer, resulting an AFM due to M-M interactions \cite{Anderson}.

\begin{figure*}
\centerline{\includegraphics[scale=1]{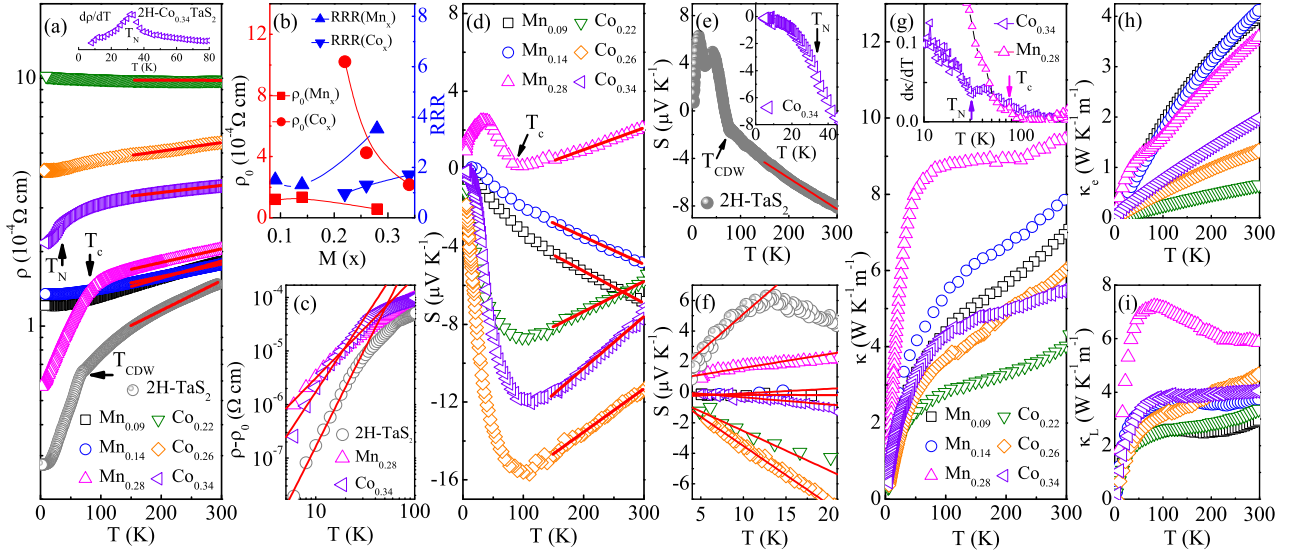}}
\caption{(Color online) (a) Temperature dependence of in-plane resistivity $\rho$(T) for 2H-M$_x$TaS$_2$ (M = Mn, Co) single crystals. Inset shows the $d\rho/dT$ for 2H-Co$_{0.34}$TaS$_2$ around $T_\textrm{N}$. (b) The evolution of residual resistivity $\rho_0$ (left axis) at $T$ = 4 K and residual-resistivity ratio RRR = $\rho_{\textrm{300 K}}/\rho_{\textrm{4 K}}$ (right axis) for 2H-M$_x$TaS$_2$ (M = Mn, Co). (c) The temperature-dependent $\rho-\rho_0$ for 2H-TaS$_2$, 2H-Mn$_{0.28}$TaS$_2$, and 2H-Co$_{0.34}$TaS$_2$ with power-law fits from 30 to 4 K (solid lines). (d) Temperature dependence of in-plane thermopower $S$(T) for 2H-M$_x$TaS$_2$ (M = Mn, Co) and (e) 2H-TaS$_2$ single crystals with linear fits from 150 to 300 K. Inset in (e) shows the $S$(T) for 2H-Co$_{0.34}$TaS$_2$. (f) The $S$(T) at low temperatures for 2H-M$_x$TaS$_2$ with linear fits. Temperature dependence of (g) total thermal conductivity $\kappa$(T), (h) electronic part $\kappa_\textrm{e}$(T), and (i) phonon part $\kappa_\textrm{L}$(T) for 2H-M$_x$TaS$_2$ (M = Mn, Co). Inset in (g) shows the $d\kappa/dT$ curves for 2H-Mn$_{0.28}$TaS$_2$ and 2H-Co$_{0.34}$TaS$_2$.}
\label{MTH}
\end{figure*}

\begin{table*}[!htb]
\caption{\label{tab} A set of parameters derived from the temperature-dependent electrical resistivity $\rho$(T), thermopower $S$(T), thermal conductivity $\kappa$(T), and heat capacity $C_p$(T) for 2H-M$_x$TaS$_2$ (M = Mn and Co) single crystals.}
\begin{ruledtabular}
\begin{tabular}{lllllllllll}
         & $x$ & $\rho_{\textrm{300 K}}$ & RRR & $S_{\textrm{300 K}}$ & $S/T$ & $\gamma$ & $q$ & $\Theta_\textrm{D}$ & $\kappa_{\textrm{300 K}}$ & $\kappa_\textrm{e}/\kappa_\textrm{L}$ \\
         & & ($10^{-4}$ $\Omega$ cm) & & ($\mu$V K$^{-1}$) & ($\mu$V K$^{-2}$) & (mJ mol$^{-1}$ K$^{-2}$) & & (K) & (W K$^{-1}$ m$^{-1}$) & (300 K)\\
  \hline
  M = Mn & 0.09 & 1.82 & 1.50 & -6.8 & -0.002(5) &&&& 6.9 & 1.35 \\
         & 0.14 & 1.78 & 1.32 & -4.8 & 0.03(1) &&&& 7.8 & 1.10 \\
         & 0.28 & 2.04 & 3.53 & 2.2 & 0.09(1) &&&& 9.4 & 0.61 \\
  \hline
  M = Co & 0.22 & 9.8 & 0.96 & -5.8 & -0.26(3) & 34.4(2) & 0.72(7) & 321(2) & 4.0 & 0.19 \\
         & 0.26 & 5.5 & 1.29 & -11.4 & -0.38(1) & 50.4(2) & 0.72(2) & 350(2) & 6.0 & 0.28 \\
         & 0.34 & 3.7 & 1.70 & -7.4 & -0.05(1) & 16.1(1) & 0.30(6) & 302(2) & 5.5 & 0.49
\end{tabular}
\end{ruledtabular}
\end{table*}

Having established the magnetic property, we proceed to investigate effects of Mn and Co intercalation on the electrical and thermal transport properties. Figure 3(a) shows the temperature dependence of in-plane resistivity $\rho$(T) for 2H-M$_x$TaS$_2$ (M = Mn, Co) single crystals. The M-intercalation removes the charge density wave (CDW) transition at 78 K for 2H-TaS$_2$ \cite{Harper}. The values of residual resistivity $\rho_0$ at 4 K and the residual resistivity ratio (RRR = $\rho_{\textrm{300 K}}/\rho_{\textrm{4 K}}$) are 2.76$\times$10$^{-5}$ $\Omega$ cm and 5.4 for 2H-TaS$_2$. All the M-intercalated samples show a metallic behavior with a larger $\rho_0$ and a smaller RRR when compared to 2H-TaS$_2$ [Table II]. The evolution of $\rho_0$ and RRR with intercalation content $x$ for 2H-M$_x$TaS$_2$ (M = Mn, Co) is plotted in Fig. 3(b). The $\rho_0$ is fairly high, presumably due to the deficiency and incomplete ordering of the Mn or Co atoms. With increasing Mn or Co content, the value of $\rho_0$ gradually decreases, agreeing with the increase of $RRR$. The mostly ordered 2H-Mn$_{0.28}$TaS$_2$ and 2H-Co$_{0.34}$TaS$_2$ might form Mn or Co superstructure in the vdW gap, calling for further electron diffraction measurement, and this ordering will minimize the electrical resistivity. The slope of $\rho$(T) shows a sharp change in magnitude at $T_\textrm{c}$ for 2H-Mn$_{0.28}$TaS$_2$, while a clear kink was observed around $T_\textrm{N}$ for 2H-Co$_{0.34}$TaS$_2$ as evidenced by the $d\rho/dT$ plot in inset of Fig. 3(a), indicating that the coupling between transport carriers in TaS$_2$ planes and local moments on intercalated-M atoms is not negligible. This effect is weaker in lower $x$ samples.

All the samples show a nearly $T$-linear behavior in the high temperature PM regime, as shown in Fig. 3(a). Upon decreasing the temperature below the magnetic transitions for 2H-Mn$_{0.28}$TaS$_2$ and 2H-Co$_{0.34}$TaS$_2$, the slope of $\rho$(T) becomes steeper, due to the decrease of spin disorder scattering \cite{Parkin2}. After subtracting the residual resistivity, the resistivity $\rho-\rho_0$ at low temperature of 2H-TaS$_2$, 2H-Mn$_{0.28}$TaS$_2$, and 2H-Co$_{0.34}$TaS$_2$ are plotted in Fig. 3(c). The power law fit ($\propto T^\alpha$) below 30 K gives $\alpha$ = 3.7(1) for  2H-TaS$_2$, 1.9(1) for 2H-Mn$_{0.28}$TaS$_2$, and 2.8(1) for 2H-Co$_{0.34}$TaS$_2$, respectively. For 2H-TaS$_2$, the low temperature $\rho-\rho_0$ is dominated by electron-phonon ($T^5$) scattering and the contribution arising from scattering of electrons by the collective excitations of the CDW ($T^2$). For 2H-Mn$_{0.28}$TaS$_2$ and 2H-Co$_{0.34}$TaS$_2$, the $\rho-\rho_0$ at low temperature varies quasi-quadratically due to probably spin-wave scattering. Above the $T_\textrm{c}$ and $T_\textrm{N}$, where the spins are disordered, we expect the magnetic scattering to be temperature-independent and it gives a linear $T$-dependence due to electron-phonon scattering [Fig. 3(a)].

Figure 3(d) shows the temperature dependence of in-plane thermopower $S$(T) for 2H-M$_x$TaS$_2$ (M = Mn, Co) single crystals. Above 150 K, the $T$-linear behavior is observed in $S$(T) similar to $\rho$(T). With decreasing temperature, the $S$(T) of 2H-TaS$_2$ changes the slope below $T_{\textrm{CDW}}$, reflecting the reconstruction of Fermi surface, and further changes its sign from negative to positive inside the CDW state featuring two peaks around 12 and 42 K [Fig. 3(e)]. Combined with the sign change of Hall coefficient across the CDW transition \cite{Michio}, multiple carriers coexist in 2H-TaS$_2$. In general, the $S$(T) is the sum of three different contributions including the diffusion term $S_{\textrm{diff}}$, the spin-dependent scattering term, and the phonon-drag term $S_{\textrm{drag}}$ due to electron-phonon coupling. The $S_{\textrm{drag}}$ term gives $\sim T^3$ for $T \ll \Theta_\textrm{D}$, $\sim T^{-1}$ for $T \gg \Theta_\textrm{D}$, and a peak structure at $\sim \Theta_\textrm{D}/5$, where $\Theta_\textrm{D}$ is the Debye temperature. The peak feature in pure 2H-TaS$_2$ may be contributed by the phonon-drag effect though the peak temperature is somehow lower than $\Theta_\textrm{D}/5 \approx$ 53(2) K \cite{Garoche}. Low Mn-intercalation obviously removes the CDW transition in $S$(T), and the sign of $S$(T) gradually changes from negative to positive with increasing Mn content $x$. For 2H-Mn$_{0.28}$TaS$_2$, the positive values of $S(T)$ in the whole temperature range indicates that hole-type carriers dominate; the thermopower changes its slope at $T_c$ and features a broad peak at 35(5) K, reflecting the reconstruction of Fermi surface passing through $T_c$ and possible phonon- or magnon-drag effect at low temperature \cite{YL1}. The thermopower values for M = Co decrease on cooling from 300 K but increase at low temperatures, after passing through minima of -8.7, -15.6, and -12.0 $\mu$V K$^{-1}$ for Co content of 0.22, 0.26, and 0.34, respectively, at a nearly same temperature 100 K. The intercept of linear fit at high temperature is rather high for a metal, which was also observed in 2H-Co$_x$NbS$_2$ \cite{Bo,Bar}. Furthermore, the $S$(T) is negative for the whole temperature range for 2H-Co$_x$TaS$_2$, in contrast to the sign of the ordinary Hall coefficient (see below), indicating a possible multi-band behavior. 2H-Co$_{0.26}$TaS$_2$ shows the largest absolute minimum value of $S$ might due to an optimized carrier concentration, mobility, and thermopower contribution from different carriers. It is visible that the $S$(T) changes its slope near $T_\textrm{N}$ for 2H-Co$_{0.34}$TaS$_2$ [inset in Fig. 3(e)], which is absent in lower $x$ samples, probably due to the fact that Co ordered moment is small.

At low temperature, the diffusive Seebeck response of Fermi liquid dominates and is also expected to be linear in $T$ [Fig. 3(f)]. Fundamentally, the thermopower is the entropy per carrier. In a simple case of free electron gas, the $S$ is given by \cite{Barnard,Miyake,Behnia},
\begin{equation}
\frac{S}{T} = \pm\frac{\pi^2}{3}\frac{k_\textrm{B}^2}{e}\frac{N(\varepsilon_\textrm{F})}{n},
\end{equation}
where $e$ is the electron charge, $k_\textrm{B}$ is the Boltzman constant, $N(\varepsilon_\textrm{F})$ is the density of states at the Fermi energy, and $n$ is the carrier concentration (the positive sign is for hole and the negative sign is for electron). The derived $S/T$ below 11 K is summarized in Table II.

Figure 3(g) shows the temperature dependence of in-plane thermal conductivity $\kappa$(T) for 2H-M$_x$TaS$_2$ (M = Mn, Co). The separate contribution of electronic $\kappa_\textrm{e}$ and phonon $\kappa_\textrm{L}$ thermal conductivities can be estimated from the Wiedemann-Franz law \cite{WH},
\begin{equation}
\frac{\kappa_\textrm{e}}{T} = \frac{\pi^2}{3}\frac{k_\textrm{B}^2}{\rho e^2},
\end{equation}
where $\rho$ is the measured resistivity. As depicted in Fig. 3(h,i), the $\kappa_\textrm{e}$ dominates in low Mn-intercalated samples, and the ratio of $\kappa_\textrm{e}/\kappa_\textrm{L}$ decreases with increasing $x$. At room temperature, the $\kappa_\textrm{e}/\kappa_\textrm{L}$ is 0.61 for 2H-Mn$_{0.28}$TaS$_2$, implying that carriers carry about two-thirds of the heat. The weak linear increase of $\kappa$(T) above 100 K is essentially attributable to electronic part, while phonon contribution is nearly temperature independent. For 2H-Co$_x$TaS$_2$, the $\kappa_\textrm{L}$ dominates in the whole temperature range, while the $\kappa_\textrm{e}$ contribution is gradually enhanced by increasing Co concentration. The rise of $\kappa_\textrm{L}$ is followed by a smooth saturation at $\sim$ 3.9 W K$^{-1}$ m$^{-1}$ above 100 K for 2H-Co$_{0.34}$TaS$_2$ [Fig. 3(i)]. The lack of the classical Umklapp maximum in $\kappa_\textrm{L}$ is probably related to a rather low value of $\kappa_\textrm{L}$ at high temperature, demonstrating a significant scattering of acoustic phonon \cite{Bar}.

\begin{figure}
\centerline{\includegraphics[scale=1]{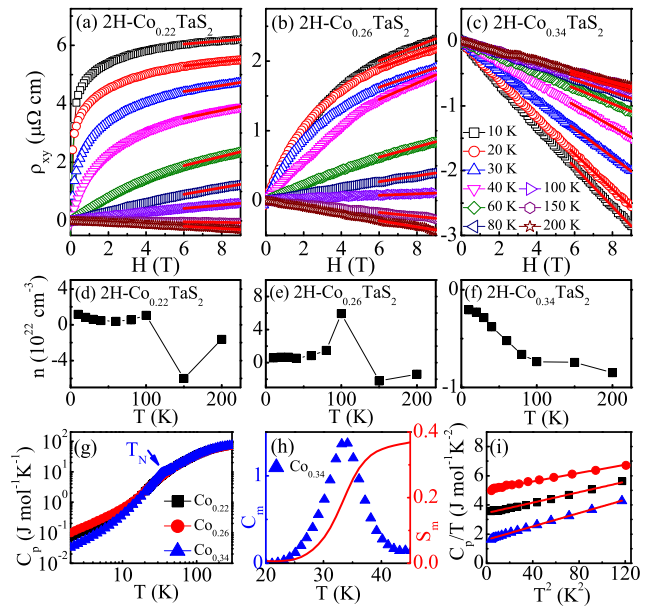}}
\caption{(Color online) (a-c) Out-of-plane field dependence of Hall resistivity $\rho_{xy}$(H) for 2H-Co$_x$TaS$_2$ at indicated temperatures with linear fits from 6 to 9 T. (d-f) The estimated carrier concentration $n$ from ordinary Hall efficient ($R_0 = 1/nq$) for 2H-Co$_x$TaS$_2$. Temperature dependence of (g) heat capacity $C_p$(T), (h) the magnetic contribution $C_m$(T) (left axis) and the derived magnetic entropy $S_m$(T) (right axis), and (i) low-temperature $C_p/T$ vs $T^2$ data fitted by $C_p/T = \gamma + \beta T^2$ for 2H-Co$_x$TaS$_2$.}
\label{MTH}
\end{figure}

To shed more light on the Co-intercalation promoted FM to AFM transition, we measured the field-dependent Hall resistivity $\rho_{xy}$(H) for 2H-Co$_x$TaS$_2$ at various temperatures [Fig. 4(a-c)]. In general, the $\rho_{xy}$(H) in ferromagnets is made up of two parts $\rho_{xy} = \rho_{xy}^\textrm{O} + \rho_{xy}^\textrm{A}$ \cite{Wang,Yan, WangY,Onoda2008}, where $\rho_{xy}^\textrm{O}$ and $\rho_{xy}^\textrm{A}$ are the ordinary and anomalous Hall resistivity, respectively. Above 100 K, the negative slope of $\rho_{xy}$(H) indicates the dominance of electron-type carries, in agreement with the $S(T)$ analysis [Fig. 3(d)], which can be accounted for by the ordinary Hall coefficient $R_0 = \rho_{xy}^\textrm{O}/H$ in the high field regime. With decreasing temperature, the slope changes sign from negative to positive for 2H-Co$_{0.22}$TaS$_2$ and 2H-Co$_{0.26}$TaS$_2$, while it keeps negative values for 2H-Co$_{0.34}$TaS$_2$ with an AFM ground state. The nonlinearity of the Hall effect at low temperatures for 2H-Co$_{0.22}$TaS$_2$ is a consequence of the interaction of the conduction electrons with the spin system. It is dominated by the extrinsic side-jump mechanism rather than extrinsic skew-scattering and intrinsic KL mechanisms \cite{YL}, where the potential field induced by impurities contributes to the anomalous group velocity. A similar feature in the intermediate Co-intercalated sample 2H-Co$_{0.34}$TaS$_2$ implies that the competition of FM and AFM, in line with the $\chi$(T) analysis [Fig. 2(g)]. In a multiband system including both electron- and hole-type carriers, the Hall coefficient gives $1/e(n_h-n_e)$ in high field regime, where $n_h$ and $n_e$ represent the hole and electron density, respectively. The estimated $n = n_h-n_e$ is plotted in Fig. 4(d-f), and the value of 10$^{22}$ cm$^{-3}$ corresponds to $\sim$ 3.5 carriers per unit cell of 2H-M$_x$TaS$_2$.

Figure 4(g) shows the temperature dependence of heat capacity $C_\textrm{p}(T)$ for 2H-Co$_x$TaS$_2$. A clear anomaly seen at $T_\textrm{N}$ corresponds well to the 3D AFM ordering for 2H-Co$_{0.34}$TaS$_2$. The high temperature $C_\textrm{p}(T)$ approaches the Dulong Petit value of 3NR $\approx$ 78 J mol$^{-1}$ K$^{-1}$ with R = 8.314 J mol$^{-1}$ K$^{-1}$. The magnetic entropy $S(T)$ = 0.37 J mol$^{-1}$ K$^{-1}$ is calculated from $S(T) = \int_0^T C_p(T,H)/TdT$ in temperature range from 20 to 45 K [Fig. 4(h)]. The low temperature data from 2 to 11 K can be well fitted by $C_\textrm{p}/T = \gamma + \beta T^2$, where the first term is the Sommerfeld electronic specific heat coefficient and the second term is low-temperature limit of lattice heat capacity [Fig. 4(i)]. The derived values of $\gamma$ as well as the Debye temperature $\Theta_\textrm{D} = (12\pi^4\textrm{NR}/5\beta)^{1/3}$, where N is the number of atoms per formula unit, are summarized in Table II.

Another expression for the electronic specific heat is:
\begin{equation}
\gamma=\frac{\pi^2k_\textrm{B}^2}{3}N(\varepsilon_\textrm{F}).
\end{equation}
Combining equations (1) and (3) yields: $S/T = \pm \gamma/ne$, where the units are V K$^{-1}$ for $S$, J K$^{-2}$ m$^{-3}$ for $\gamma$, and m$^{-3}$ for $n$, respectively. In order to compare different materials, it is common to express $\gamma $ in unit J mol$^{-1}$ K$^{-2}$. Then we define a dimensionless quantity
\begin{equation}
q=\frac{S}{T}\frac{N_\textrm{A}e}{\gamma},
\end{equation}
where $N_\textrm{A}$ is the Avogadro number. The constant $N_\textrm{A}e$ = 9.6$\times$10$^{4}$ C mol$^{-1}$ is also called the Faraday number. The $q$ gives the number of carriers per formula unit (proportional to $1/n$) \cite{Behnia}. The obtained $q$ = 0.30(6) for 2H-Co$_{0.34}$TaS$_2$ indicates about 3.3(7) electrons per formula unit within the Boltzmann framework, close to the value of 3.5 carriers estimated from the ordinary Hall coefficient. The $q$ $\sim$ 0.72 is about two times larger for 2H-Co$_{0.26}$TaS$_2$ and 2H-Co$_{0.22}$TaS$_2$, which is consistent with the electronic doping produced by Co-intercalation.

\section{CONCLUSIONS}

In summary, we systematically studied the evolution of magnetic, electrical and thermal transport properties for a series of M-intercalated 2H-M$_x$TaS$_2$ (M = Mn, Co) vdW magnets. The magnetic transition corresponds well to a kink in resistivity, a weak anomaly in thermal conductivity, as well as a slope change in thermopower for 2H-Mn$_{0.28}$TaS$_2$ and 2H-Co$_{0.34}$TaS$_2$, but is weaker for lower $x$ crystals. Thermopower at low temperatures can be well described by a diffusive thermoelectric response model implying the dominant electronic contribution and small electron-phonon coupling. Carrier concentration analysis indicates that intercalated Co atoms produce electronic doping via hybridization with atoms around vdW gap. It is also of high interest to explore the thickness-dependent properties of 2H-M$_x$TaS$_2$ (M = Mn, Co) at the 2D limit down to a monolayer in future studies as well as the intercalated metal monolayer excitations for material properties and developing applications \cite{Fan}.

\section*{Acknowledgements}

Work at BNL is supported by the Office of Basic Energy Sciences, Materials Sciences and Engineering Division, U.S. DOE under Contract No. DE-SC0012704. This research used resources of the Center for Functional Nanomaterials, which is a U.S. DOE Office of Science Facility, at BNL under Contract No. DE-SC0012704. Y.L. acknowledges a Director$^\prime$s Postdoctoral Fellowship through the Laboratory Directed Research and Development program of the Los Alamos National Laboratory. E.D.B. was supported by the Office of Basic Energy Sciences, Materials Sciences and Engineering Division, U.S. DOE under the ``Quantum Fluctuations in Narrow-Band Systems."\\

$\dag$ yuliu@lanl.gov
$\ddag$ petrovic@bnl.gov


\begin{references}
\bibitem{Geim} A. K. Geim and I. V. Grigorieva, Van der Waals heterostructures, Nature \textbf{499}, 419 (2013).
\bibitem{Novoselov} K. S. Novoselov, A. Mishchenko, A. Carvalho, and A. H. Castro Neto, 2D materials and van der Waals heterostructures, Science \textbf{353}, 461 (2016).
\bibitem{Lee} J. Lee, S. Lee, J. H. Ryoo, S. Kang, T. Y. Kim, P. Kim, C. Park, J. Park, and H. Cheong, Ising-Type Magnetic Ordering in Atomically Thin FePS$_3$, Nano Lett. \textbf{16}, 7433 (2016).
\bibitem{Huang} B. Huang, G. Clark, E. Navarro-Moratalla, D. R. Klein, R. Cheng, K. L. Seyler, D. Zhong, E. Schmidgall, M. A. McGuire, D. H. Cobden, W. Yao, D. Xiao, P. Jarillo-Herrero, and X. D. Xu, Layer-dependent ferromagnetism in a van der Waals crystal down to the monolayer limit, Nature \textbf{546}, 270 (2017).
\bibitem{Gong} C. Gong, L. Li, Z. L. Li, H. W. Ji, A. Stern, Y. Xia, T. Cao, W. Bao, C. Z. Wang, Y. Wang, Z. Q. Qiu, R. J. Cava, S. G. Louie, J. Xia, and X. Zhang, Discovery of intrinsic ferromagnetism in two-dimensional van der Waals crystals, Nature \textbf{546}, 265 (2017).
\bibitem{Deng} Y. J. Deng, Y. J. Yu, Y. C. Song, J. Z. Zhang, N. Z. Wang, Z. Y. Sun, Y. F. Yi, Y. Z. Wu, S. W. Wu, J. Y. Zhu, J. Wang, X. H. Chen, and Y. B. Zhang, Gate-tunable room-temperature ferromagnetism in two-dimensional Fe$_3$GeTe$_2$, Nature \textbf{563}, 94 (2018).
\bibitem{Bonilla} M. Bonilla, S. Kolekar, Y. Ma, H. C. Diaz, V. Kalappattil, R. Das, T. Eggers, H. R. Gutierrez, M. Phan, and M. Batzill, Strong room-temperature ferromagnetism in VSe$_2$ monolayers on van der Waals substrates, Nat. Nanotechnol. \textbf{13} 289 (2018).
\bibitem{Hara} D. J. O'Hara, T. Zhu, A. H. Trout, A. S. Ahmed, Y. K. Luo, C. H. Lee, M. R. Brenner, S. Rajan, J. A. Gupta, D. W. McComb, and R. K. Kawakami, Room Temperature Intrinsic Ferromagnetism in Epitaxial Manganese Selenide Films in the Monolayer Limit, Nano Lett. \textbf{18}, 3125 (2018).
\bibitem{Parkin} S. S. P. Parkin and R. H. Friend, $3d$ transition-metal intercalates of the niobium and tantalum dichalcogenides. I. Magnetic properties, Philos. Mag. B \textbf{41}, 65 (1980).
\bibitem{Friend0} R. H. Friend and A. D. Yoffe, Electronic properties of intercalation complexes of the transition metal dichalcogenides, Advances in Phys. \textbf{36}, 1 (1987).
\bibitem{Hinode} H. Hinode, T. Ohtani, and M. Wakihara, Homogeneity Range and Some Physical Properties of Intercalation Compounds of Mn$_x$TaS$_2$, J. Solid State Chem. \textbf{114}, 1 (1995).
\bibitem{Onuki} Y. Onuki, K. Ina, T. Hirai, and T. Komatsubara, Magnetic properties of intercalation compoumd: Mn$_{1/4}$MX$_2$, J. Phys. Soc. Jpn. \textbf{55}, 347 (1986).
\bibitem{Cussen} L. D. Cussen, E. A. Marseglia, D. M. Paul, and B. D. Rainford, Magnon dispersion in Mn$_{1/4}$TaS$_2$, Physica B: Condens. Matter \textbf{156-157}, 712 (1989)
\bibitem{Li} L. J. Li, W. J. Lu, X. D. Zhu, Z. R. Zhang, W. H. Song, and Y. P. Sun, Influence of the low Mn intercalation on magnetic and electronic properties of 2H-TaS$_2$ single crystals, J. Magn. Magn. Mater. \textbf{323}, 2536 (2011).
\bibitem{Shand} P. M. Shand, C. Cooling, C. Mellinger, J. J. Danker, T. E. Kidd, K. R. Boyle, and L. H. Strauss, Magnetic states in nanostructured manganese-intercalated TaS$_2$, J. Magn. Magn. Mater. \textbf{382}, 49 (2015).
\bibitem{Zhang1} H. Zhang, W. Wei, G. Zheng, J. Lu, M. Wu, X. Zhu, J. Tang, W. Ning, Y. Han, L. Lin, J. Yang, W. Gao, Y. Qin, and M. Tian, Electrical and anisotropic magnetic properties in layered Mn$_{1/3}$TaS$_2$ crystals, Appl. Phys. Lett. \textbf{113}, 072402 (2018).
\bibitem{YL1} Y. Liu, Z. Hu, E. Stavitski, K. Attenkofer, and C. Petrovic, Three-dimensional ferromagnetism and magnetotransport in van der Waals Mn-intercalated tantalum disufide, Phys. Rev. B \textbf{103}, 144432 (2021).
\bibitem{Morosan} E. Morosan, H. W. Zandbergen, L. Li, M. Lee, J. G. Checkelsky, M. Heinrich, T. Siegrist, N. P. Ong, and R. J. Cava, Sharp switching of the magnetization in Fe$_{1/4}$TaS$_2$, Phys. Rev. B \textbf{75}, 104401 (2007).
\bibitem{Hardy} W. J. Hardy, C.-W. Chen, A. Marcinkova, H. Ji, J. Sinova, D. Natelson, and E. Morosan, Very large magnetoresistance in Fe$_{0.28}$TaS$_2$ single crystals, Phys. Rev. B \textbf{91}, 054426 (2015).
\bibitem{Chen} C.-W. Chen, S. Chikara, V. S. Zapf, and E. Morosan, Correlations of crystallographic defects and anisotropy with magnetotransport properties in Fe$_x$TaS$_2$ single crystals (0.23 $\leq x \leq$ 0.35), Phys. Rev. B \textbf{94}, 054406 (2016).
\bibitem{Cai} R. Cai, W. Xing, H. Zhou, B. Li, Y. Chen, Y. Yao, Y. Ma, X. C. Xie, S. Jia, and W. Han, Anomalous Hall effect mechanisms in the quasi-two-dimensional van der Waals ferromagnet Fe$_{0.29}$TaS$_2$, Phys. Rev. B \textbf{100}, 054430 (2019).
\bibitem{Zhang} C. Zhang, Y. Yuan, M. Wang, P. Li, J. Zhang, Y. Wen, S. Zhou, and X. X. Zhang, Critical behavior of intercalated quasi-van der Waals ferromagnet Fe$_{0.26}$TaS$_2$, Phys. Rev. Mater. \textbf{3}, 114403 (2019).
\bibitem{YL} Y. Liu, Z. Hu, E. Stavitski, K. Attenkofer, and C. Petrovic, Magnetic critical behavior and anomalous Hall effect in 2H-Co$_{0.22}$TaS$_2$ single crystals, Phys. Rev. Research \textbf{3}, 023181 (2021).
\bibitem{VanLaar} B. Van Laar, H. M. Rietveld and D. J. W. Ijdo, Magnetic and crystallographic structures of Me$_x$NbS$_2$ and Me$_x$TaS$_2$, J. Solid State Chem. \textbf{3}, 154 (1971).
\bibitem{LiLijun} L. Li, J. Zhang, G. Myeong, W. Shin, H. Lim, B. Kim, S. Kim, T. Jin, S. Cavill, B. S. Kim, C. Kim, J. Lischner, A. Ferreira, and S. Cho, Gate-Tunable Reversible Rashba-Edelstein Effect in a Few-Layer Graphene/2H-TaS$_2$ Heterostructure at Room Temperature, ACS Nano \textbf{14}, 5251 (2020).
\bibitem{Zhao} B. C. Zhao, Y. Q. Ma, W. H. Song, and Y. P. Sun, Magnetization steps in the phase separated manganite La$_{0.275}$Pr$_{0.35}$Ca$_{0.375}$MnO$_3$, Phys. Lett. A \textbf{354}, 472 (2006).
\bibitem{Bongers} P. F. Bongers, C. F. van Bruggen, J. Koopstra, W. P. F. A. M. Omloo, G. A. Wiegers, and F. Jellinek, Structures and magnetic properties of some metal (I) chromium (III) sulfides and selenides, J. Phys. Chem. Solids \textbf{29}, 977 (1968).
\bibitem{Parkin0} S. S. P. Parkin, E. A. Marseglia, and P. J. Brown, Magnetisation density distribution in Mn$_{1/4}$TaS$_2$: observation of conduction electron spin polarisation, J. Phys. C.: Solid State Phys. \textbf{16}, 2749 (1983).
\bibitem{Wohlfarth} E. P. Wohlfarth, Magnetic properties of crystalline and amorphous alloys: A systematic discussion based on the Rhodes-Wohlfarth plot, J. Magn. Magn. Mater., \textbf{7}, 113 (1978).
\bibitem{Moriya} T. Moriya, Recent progress in the theory of itinerant electron magnetism, J. Magn. Magn. Mater., \textbf{14}, 1 (1979).
\bibitem{Hechang} S. Tian, J. Zhang, C. Li, T. Ying, S. Li, X. Zhang, K. Liu, and H. Lei, Ferromagnetic van der Waals crystals VI$-3$, J. Am. Chem. Soc. \textbf{141}, 5326 (2019).
\bibitem{Parkin1} S. S. P. Parkin, E. A. Marseglia, and P. J. Brown, Magnetic structure of Co$_{1/3}$NbS$_2$ and Co$_{1/3}$TaS$_2$, J. Phys. C.: Solid State Phys. \textbf{16}, 2765 (1983).
\bibitem{Choe} J. Choe, K. Lee, C. L. Huang, N. Trivedi, and E. Morosan, Magnetotransport in Fe-intercalated $T$S$_2$: comparison between $T$ = Ti and Ta, Phys. Rev. B \textbf{99}, 064420 (2019).
\bibitem{Dijkstra} J. Dijkstra, P. J. Zijlema, C. F. van Bruggen, C. Haas, and R. A. de Groot, Band-structure calculations of Fe$_{1/3}$TaS$_2$ and Mn$_{1/3}$TaS$_2$, and transport and magnetic properties of Fe$_{0.28}$TaS$_2$, J. Phys.: Condens. Matter \textbf{1}, 6363 (1989).
\bibitem{Aristov} D. N. Aristov, Indirect RKKY interaction in any dimensionality, Phys. Rev. B \textbf{55}, 8064 (1997).
\bibitem{Yosida} K. Yosida, Magnetic Properties of Cu-Mn Alloys, Phys. Rev. \textbf{106}, 893 (1957).
\bibitem{Fukuma} Y. Fukuma, H. Asada, S. Miyawaki, T. Koyanagi, S. Senba, K. Goto, and H. Sato, Carrier-induced ferromagnetism in Ge$_{0.92}$Mn$_{0.08}$Te epilayers with a Curie temperature up to 190 K, Appl. Phys. Lett. \textbf{93}, 252502 (2008).
\bibitem{YamasakiY} Y. Yamasaki, R. Moriya, M. Arai, S. Masubuchi, S. Pyon, T. Tamegai, K. Ueno and T. Machida, Exfoliation and van der Waals heterostructure assembly of intercalated ferromagnet Cr$_{1/3}$TaS$_2$, 2D Mater. \textbf{4}, 041007 (2017).
\bibitem{NiuJ} J. Niu, W. Zhang, Z. Li, S. Yang, D. Yan, S. Chen, Z. Zhang, Y. Zhang, X. Ren, P. Gao, Y. Shi, D. Yu and X. Wu, Intercalation of van der Waals layered materials: A route towards engineering of electron correlation, Chin. Phys. B \textbf{29}, 097104 (2020).
\bibitem{Narita} H. Narita, H. Ikuta, H. Hinode, T. Uchida, T. Ohtani, and M. Wakihara, Preparation and physical properties of Fe$_x$TaS$_2$ ($0.15\leq x \leq 0.50 $) compounds, J. Solid State Chem. \textbf{108}, 148 (1994).
\bibitem{Anderson} P. W. Anderson, Theory of magnetic exchange interactions: exchange in insulators and semiconductors, Solid State Physics \textbf{14}, 99 (1963).
\bibitem{Harper} J. M. Harper, T. E. Geballe and F. J. Disalvo, Thermal properties of layered transition-metal dichalcogenides at charge-density-wave transitions, Phys. Rev. B \textbf{15} 2943 (1977).
\bibitem{Parkin2} S. S. P. Parkin and R. H. Friend, Magnetic and transport properties of 3d transition metal intercalates of some group Va transition metal dichalcogenides, Physica B+C \textbf{99}, 219 (1980).
\bibitem{Michio} M. Naito and S. Tanaka, Electrical transport properties in 2H-TaS$_2$, -NbSe$_2$, -TaS$_2$ and -TaSe$_2$, J. Phys. Soc. Jpn. \textbf{51}, 219 (1981).
\bibitem{Garoche} P. Garoche, P. Manuel, J. J. Veyssi\'{e}, and P. Molini\'{e}, Dynamic measurements of the low-temperature specific heat of 2H-TaS$_2$ single crystals in magnetic fields, J. Low Temp. \textbf{30}, 323 (1978).
\bibitem{Bo} H. J. M. Bouwmeester, A. van der Lee, S. van Smaalen, and G. A. Wiegers, Order-disorder transition in silver-intercalated niobium disulfide compounds, Phys. Rev. B \textbf{43}, 9431 (1991).
\bibitem{Bar} N. Bari\v{s}i\'{c}, I. Smiljani\'{c}, P. Pop\v{c}evi\'{c}, A. Bilu\v{s}i\'{c}, E. Tuti\v{s}, A. Smontara, H. Berge, J. Ja\'{c}imovi\'{c}, O. Yuli, and L. Forr\'{o}, High-pressure study of transport properties in Co$_{0.33}$NbS$_2$, Phys. Rev. B \textbf{84}, 075157 (2011).
\bibitem{Barnard} R. D. Barnard, \textit{Thermoelectricity in Metals and Alloys} (Taylor \& Francis, London, 1972).
\bibitem{Miyake} K. Miyake and H. Kohno, Theory of quasi-universal ratio of Seebeck coefficient to specific heat in zero-temperature limit in correlated metals, J. Phys. Soc. Jpn. \textbf{74}, 254 (2005).
\bibitem{Behnia} K. Behnia, D. Jaccard and J. Flouquet, On the thermoelectricity of correlated electrons in the zero-temperature limit, J. Phys.: Condens. Matter. \textbf{16}, 5187 (2004).
\bibitem{WH} R. Franz and G. Wiedemann, Ueber die W\"{a}rme-Leitungsf\"{a}higkeit der Metalle, Annalen der Physik \textbf{165}, 497 (1853).
\bibitem{Wang} Q. Wang, S. S. Sun, X. Zhang, F. Pang, and H. C. Lei, Anomalous Hall effect in a ferromagnetic Fe$_3$Sn$_2$ single crystal with a geometrically frustrated Fe bilayer kagome lattice, Phys. Rev. B \textbf{94}, 075135 (2016).
\bibitem{Yan} J. Yan, X. Luo, G. T. Lin, F. C. Chen, J. J. Gao, Y. Sun, L. Hu, P. Tong, W. H. Song, Z. G. Sheng, W. J. Lu, X. B. Zhu, and Y. P. Sun, Anomalous Hall effect of the quasi-two-dimensional weak itinerant ferromagnet Cr$_{4.14}$Te$_8$, Europhys. Lett. \textbf{124}, 67005 (2018).
\bibitem{WangY} Y. H. Wang, C. Xian, J. Wang, B. J. Liu, L. S. Ling, L. Zhang, L. Cao, Z. Qu, and Y. M. Xiong, Anisotropic anomalous Hall effect in triangular itinerant ferromagnet Fe$_3$GeTe$_2$, Phys. Rev. B \textbf{96}, 134428 (2017).
\bibitem{Onoda2008} S. Onoda, N. Sugimoto, and N. Nagaosa, Quantum transport theory of anomalous electric, thermoelectric, and thermal Hall effects in ferromagnets, Phys. Rev. B \textbf{77}, 165103 (2008).
\bibitem{Fan} S. Fan, S. Neal, C. Won, J. Kim, D. Sapkota, F. Huang, J. Yang, D. G. Mandrus, S. -W. Cheong, J. T. Haraldsen, and J. L. Musfeldt, Excitations of intercalated metal monolayers in transition metal dichalcogenides, Nano Letters \textbf{19}, 99 (2021).
\end{references}
\end{document}